\documentclass[10pt,a4paper,twoside]{article}
\usepackage{epsfig}
\usepackage{baltlat6}
\usepackage{array}
\usepackage{here}
\usepackage[cp1251]{inputenc}
\usepackage[russian]{babel}

\begin{document}
\ \
\vspace{0.5mm}
\setcounter{page}{1}
\vspace{8mm}

\titlehead{Baltic Astronomy, vol.\,??, ??--??, 2011}

\titleb{SCORPIO AT THE  6-M TELESCOPE: CURRENT STATE AND PERSPECTIVES FOR SPECTROSCOPY OF GALACTIC AND EXTRAGALACTIC OBJECTS}

\begin{authorl}
\authorb{V.L. Afanasiev}{} and
\authorb{A.V Moiseev}{}
\end{authorl}

\begin{addressl}
\addressb{}{Special Astrophysical Observatory of RAS, Nizhnij Arkhyz, Russia 369167; vafan@sao.ru, moisav@gmail.com}
\end{addressl}

\submitb{Received: 2011 June 10; accepted: 2011 ??}

\begin{summary} A significant part of observations by Russian 6-m telescope  is carried out using SCORPIO multi-mode focal reducer.   A lot of scientific data  have been collected using observations in  direct imaging, slit spectroscopy and Fabry-P\'{e}rot interferometry modes during the past ten years. Some results of these observations are considered in this review.  We are also  present a short description of a new generation instrument named SCORPIO-2.
\end{summary}


\resthead{SCORPIO at the  6-m telescope}
{V. Afanasiev, A. Moiseev}

\sectionb{1}{INTRODUCTION}

In the middle of the last century Georg Court\'{e}s  (1960) suggested and realized the idea of a focal reducer. In addition to increasing the  field-of-view of a  large optical telescope and correction of the off-axis primary mirror aberration, a focal reducer makes it possible  to have a multi-mode instrument, because it becomes possible to install dispersing elements  in a output pupil between the collimator and the camera, which turns the direct imaging system into an universal spectrograph.  The first prototype of the device   designed for  spectroscopy and photometry of faint extended objects was an  EFOSC camera at the 3.6-m ESO telescope (Buzzoni et al. 1984). Now  a lot of multimode low-resolution spectrographs are used at middle-size and large telescopes.  SCORPIO (Spectra Camera	with	 Optical	 Reducer for Photometric and Interferometric Observations) have worked at the primary focus of the 6-m SAO RAS telescopes since Sep. 2000. In the paper by Afanasiev \& Moiseev (2005) we gave a short description o the device,  while technical  details are considered in Afanasiev et al. (2005).   See also current information  presented on  the SCORPIO web-page (see the footnote below). Today  it is   the most frequently used facility, that has been  employed   half of  observation  time at the 6-m telescope (Figure~1). In this review we  consider briefly  some scientific results obtained  with the use of SCORPIO  and also our current work to modify and  improve of a technique of a faint object spectroscopy by  the SAO RAS telescope.

\begin{figure}[!tH]
\vbox{
\centerline{
\psfig{figure=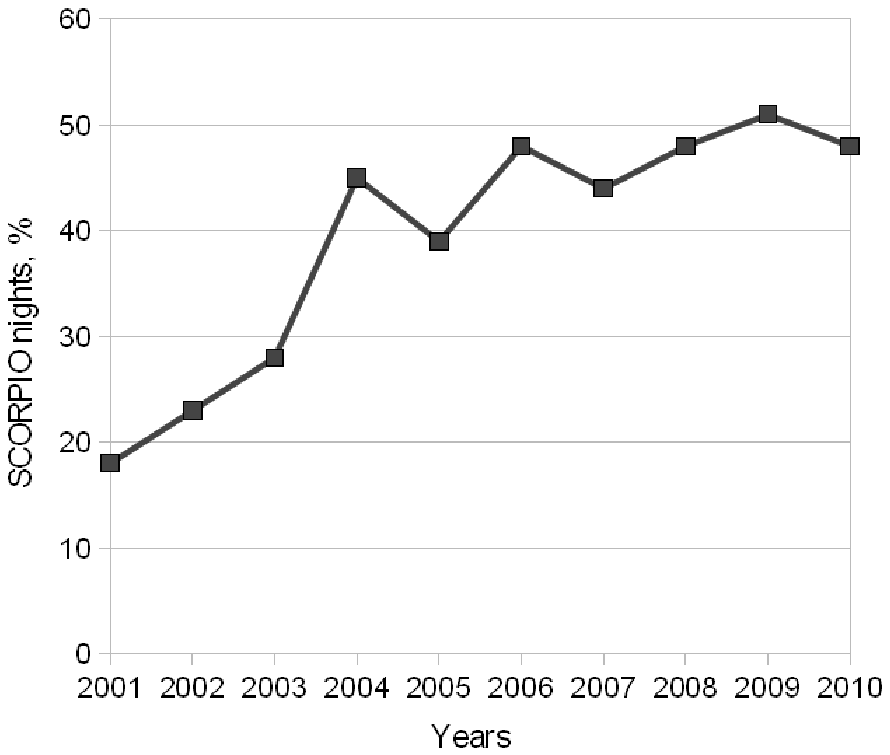,height=45mm,angle=0,clip=}
\psfig{figure=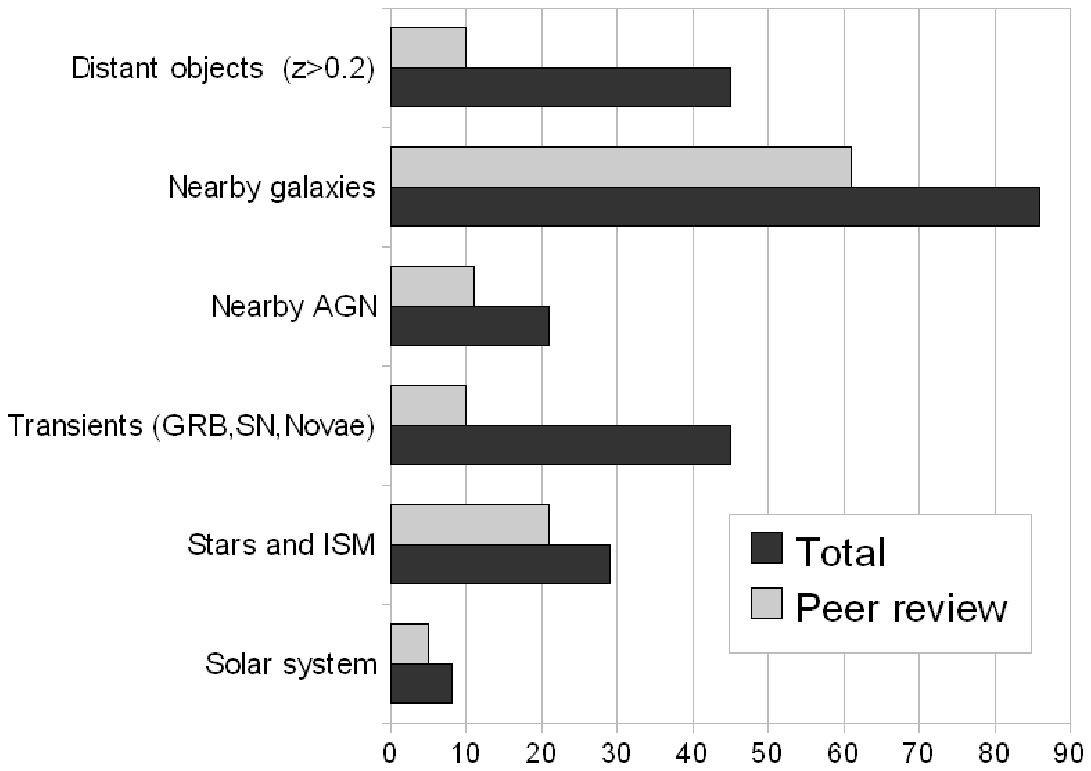,height=45mm,angle=0,clip=}
}
\vspace{1mm}
\captionb{1}
{The percentage of the calendar time distributed by 6-m telescope Program Committee for SCORPIO observations (left). Distribution of publications based on SCORPIO observations in 2001-2011 between different topics (right): the total amount (black) and  in peer-review journals only (gray).}
}
\vskip-3mm
\end{figure}

\sectionb{2}{SCORPIO OBSERVING MODES}

The  multimode focal reducer allows various spectroscopic and photometric observations to be performed within 6 arcmin field-of-view (see, Figure~2).  The list of observing modes is as  follows.

\begin{itemize}
\item Direct imaging in broad-band $UBVR_CI_C$ filters;  medium-  and narrow-band interferometric filters.
\item Long-slit spectroscopy with volume phase holographic gratings (VPHGs).
\item Slitless spectroscopy for observations of spectrophotometric standard stars.
\item Multi-slit unit for  spectroscopy with 16 movable slits ($18''$ in each height).
\item Spectropolarimetry using the analyzer based on a Savar plate  (see Afanasiev \& Moiseev (2005) for details of the data analysis).
\item 3D spectroscopy with a scanning Fabry-P\'{e}rot interferometer (FPI). The realization  of this technique in SCORPIO as well as a data reduction are  described in Moiseev et al. (2002) and in Moiseev \& Egorov (2008)
 \end{itemize}

We can  change the modes during the same  night of observation, however some restrictions exist, for instance, it is impossible to switch between long-slit and multi-slit modes. The quick switch between the main modes  (long-slit/imaging, FPI/imaging, etc.) allows an observer to  choose targets  that  match best  the current atmospheric conditions (i.e. seeing and transparency). That is very important in  case of unstable weather around the 6-m telescope site.

\sectionb{3}{SCIENTIFIC RESULTS}

According to the ADS database for   the   2001  to  June  2011 period, the data obtained with SCORPIO were presented in 215 publications, including peer-review articles, conference proceedings, telegrams, etc\footnote{The updated list of publications is available at the SCORPIO web-page: \verb"http://www.sao.ru/hq/lsfvo/devices/scorpio/scorpio.html"}.  They  have been  cited more than 1000 times. Results of these observations were used   in more than fifteen Ph.D. theses. Figure~1 (right) shows the distribution of these publication  between different astrophysics topics, which reflects the main  interest  of the astronomers who requested observing time with SCORPIO. \textit{It would be quite impossible in a short paper to give a full review of all published results. Therefore, below we will consider only certain works selected according to our preferences in order to show a large range of tasks and methods.}

\subsectionb{3.1}{Solar System}

The activity of a number  of distant comets   was investigated using the  photometric and spectroscopic  observations with SCORPIO -- see, for instance,  Korsun et al. (2008, 2010). The origin of the activity at the distance larger than 5~AU is a puzzle, however,  molecular emissions were found in some objects. For example,   in   C/2002 VQ94 (LINEAR)  emission bands of CO$^+$ and N$_2^+$ were detected  at a record heliocentric distance  -- 7.3 AU. It  is an evidence of the fact that they had been formed in the outer regions of the Solar System or in a pre-solar interstellar cloud in a low temperature ($T<25$ K) environment.

An interesting result was obtained by Afanasiev et al. (2007) who  recorded the spectrum of a faint meteor during the observations of the spectra of galaxies with the multi-slit unit.  The velocity of the entry  of the meteor body into the Earth’s atmosphere estimated from the emission lines  line-of-sight velocity  is about   300 km/s. Based on this results authors supposed that this meteor particle is likely to be of an extragalactic origin.

\subsectionb{3.2}{Stars and Interstellar Medium}

Thanks to high optics transparence and the large diameter of the telescope  mirror, SCORPIO is widely used for  snapshot and monitoring observations of faint transient objects in the frameworks of several  programmes with aims to study  spectral evolution of Novae (see for instance, Fabrika et al.  2009) and Supernovae stars, including distant core-collapse  SNe probably associated with gamma-ray bursts (Moskivitin et al.  2010). SCORPIO data provide a spectral confirmation  for  newly discovered massive evolved stars (WR, LBV) in our Galaxy (Gvaramadze et al.  2009) and other nearby galaxies: M33 (Valeev et al.  2009) and DDO68. In the latter case Pustilnik et al. (2008) have discovered a luminous blue variable  from the transient event in the spectra of HII region of DDO68.

Observations with a scanning FPI makes it possible  to study the structure of desired spectral lines (H$\alpha$, [O\,III], [SII]) simultaneously in a  large (6 arcmin) field-of-view. It provides rich opportunities for investigating   the emission-line kinematics of the ionized gas in interplay between stars and the surrounding medium. The examples in our Galaxy are the study of jets and emission knots ejected from young stellar objects (Movsessian et al. 2007, 2009),  kinematics of bow shock fronts in pulsar-wind nebula   CTB 80 (Lozinskaya et al.  2005), and   supersonic motions optical filament in the  radio nebula W50 around the   microquasar SS433 (Abolmasov et al.  2010). The related object -- a nebular complex associated with the ultraluminous X-ray sources  in the dwarf galaxy  HoIX was also studied with the SCORPIO/FPI (Abolmasov \& Moiseev 2008). Based on the SCORPIO long-slit and FPI data Lozinskaya \& Moiseev   (2007) have presented  evidences that an explosion of a very massive star (Hypernova)  seems to be a more plausible mechanism of formation of the synchrotron superbubble in IC10 galaxy compared with the earlier proposed model of multiple supernova explosions. This work is a part of a series of papers  aimed at investigating the kinematics of shells and bubbles around  star formation regions in nearby dwarf galaxies. A good illustration is  the   IC1613 galaxy where Lozinskaya et al. (2003) have estimated the expansion velocities of multiple gaseous  shells using  spatial-resolved kinematic data for ionized (H$\alpha$, SCORPIO/FPI) and neutral (21 cm, VLA) interstellar medium.

\begin{figure}[!tH]
\vbox{
\centerline{
\psfig{figure=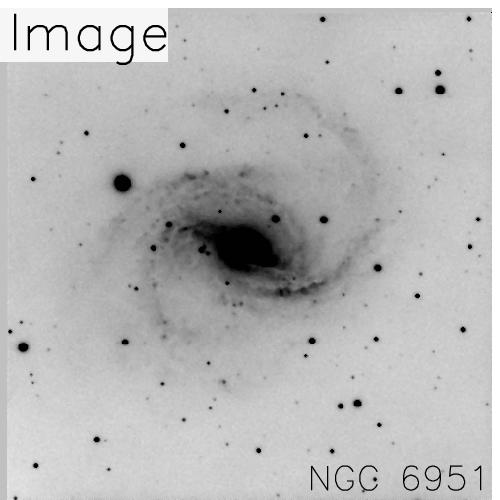,height=40mm,angle=0}
\psfig{figure=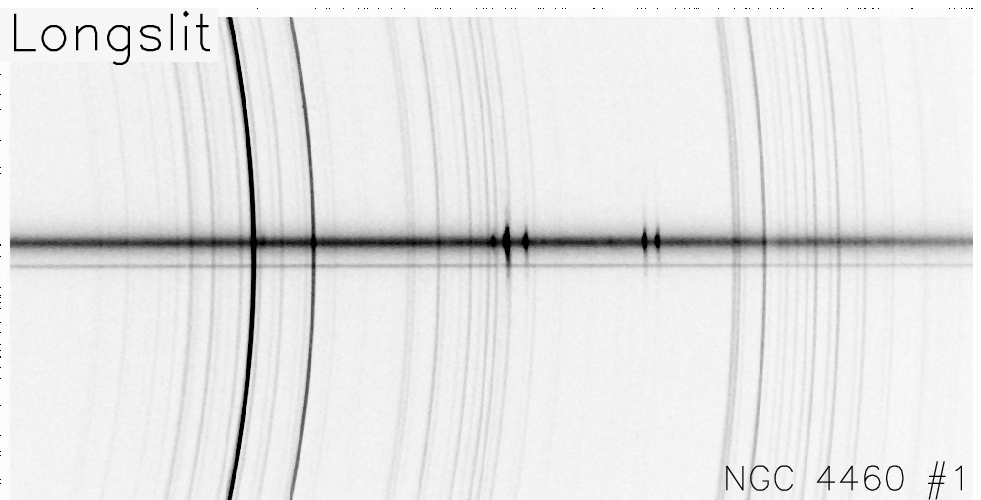,height=40mm,angle=0}
}
\centerline{
\psfig{figure=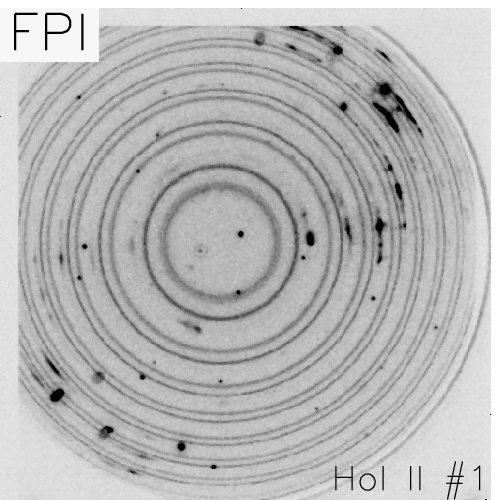,height=40mm,angle=0}
\psfig{figure=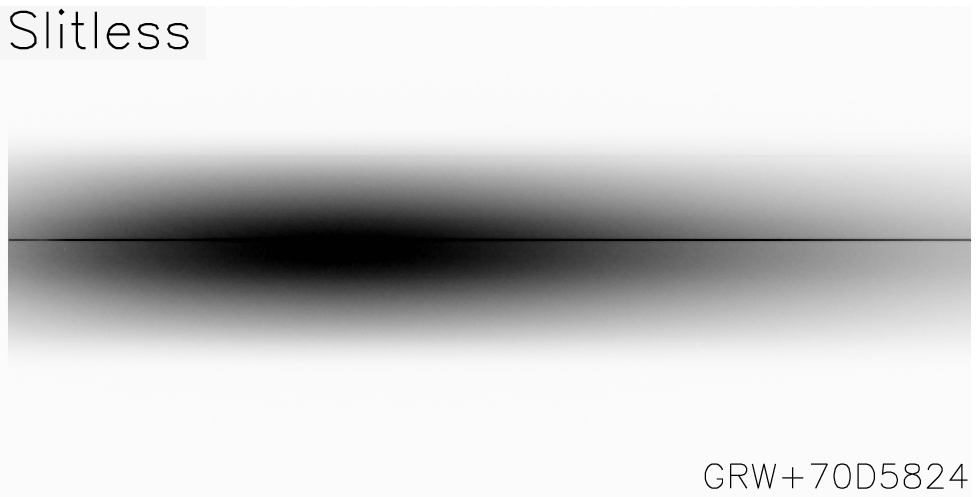,height=40mm,angle=0}
}
\centerline{
\psfig{figure=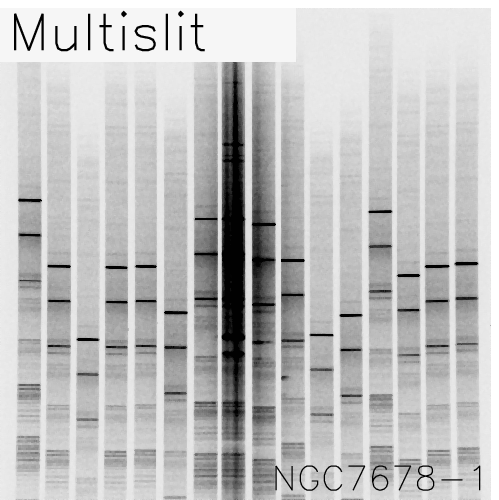,height=40mm,angle=0}
\psfig{figure=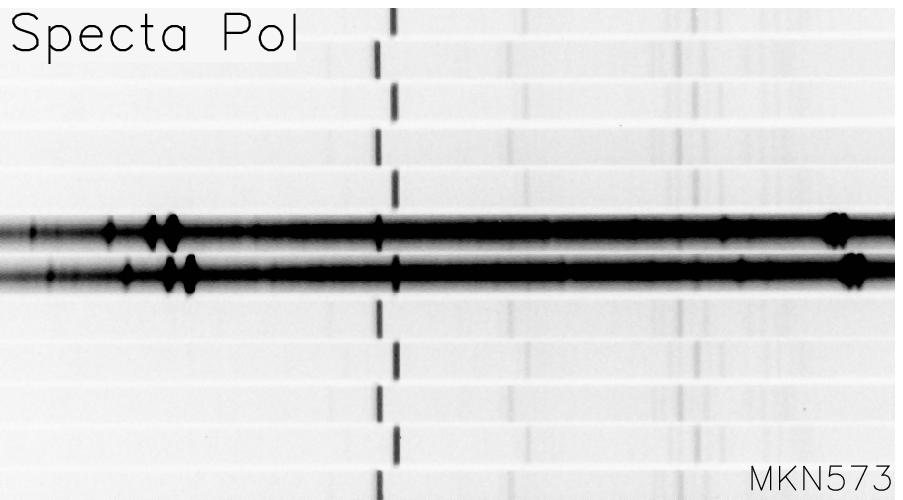,height=40mm,angle=0}
}

\vspace{1mm}
\captionb{2}{Examples of data frames for different SCORPIO modes}
}
\end{figure}

\subsectionb{3.3}{Nearby Galaxies}

Figure~1 shows that  most of SCORPIO publications are related to nearby  galaxies.   The H$\alpha$ images for  a significant part of all galaxies in the Local Volume (within 10 Mpc) were obtained during a general  imaging survey with SCORPIO. Measured H$\alpha$ fluxes were  used to derive the total star formation rate  density in the
Local Universe --  $0.019\pm0.003\,M_\odot\,yr^{-1}\,Mpc^{-3}$ (Karachentsev \& Kaisin 2010). The SCORPIO long-slit spectra were used to study  the stellar population  in two dE/dSph members of the nearby M81 group of galaxies
(Makarova et al. 2010), whereas Chilingarian et al. (2009) used the  multi-slit unit  for the following-up spectroscopy of new discovered compact elliptical  galaxies in order to investigate their origin and stellar population properties.

Above we already  discussed the ionized gas  properties  in the nearby dwarf galaxies. Using the SCORPIO/FPI observations Mart\'{\i}nez-Delgado et al. (2007) have mapped the regions of supersonic gas motions in more distant blue compact galaxies.  They offered  kinematic diagnostic diagrams that provide a possibility to infer from FPI data
the magnitude of the star formation activity in galaxies even if they are not spatially resolved. The  spectrophotometric observations conducted with the use of SCORPIO allow one to estimate the oxygen abundance in HII regions of extremely metal-deficient galaxies (see Pustilnik et al. (2010) and reference therein). A detailed analysis of   ionized gas morphology and kinematics in nine  such galaxies  shows  the important role of recent interactions and mergers in the triggering of their star formation (Moiseev et al. 2010).

The ionized gas velocity fields derived from SCORPIO/FPI data cubes reveal a complex kinematic picture in the disc of spiral galaxies caused by internal  (seculiar evolution) and external (merging, gas accretion) effects: inflow steaming motions in bars,  polar discs and rings, circumnuclear counter-rotated component (see previous review in  Moiseev 2007).  FPI kinematic mapping is very helpful in study of structure and dynamics of peculiar galaxies: colliding ring and  polar ring galaxies (see references in the review by  Moiseev \&  Bizyaev 2009). Polar rings  are an interesting example of peculiar systems that reveal outer rings or discs, rotating in the plane approximately perpendicular to the disc of the main galaxy. The recent progress in the study of polar rings with SCORPIO was presented in the paper by Brosch  et al. (2010) who found the most distant kinematically  confirmed polar ring ($z = 0.06$). Here an early-type central galaxy is surrounded by a giant (with a diameter of over 48 kpc) ring of young stars and clouds of ionized
gas, inclined at a steep angle to the stellar disc. In contrast to this large-scale structures, Moiseev (2010) has described the smallest ($r<2 $ kpc) polar gaseous discs in blue compact dwarf galaxies. The possible formation mechanism for these discs are merging or accretion of external gas clouds with a specific direction of an orbital momentum. It was also suggested by Sil'chenko et al. (2011) who studied the stellar population and kinematics properties in NGC7217 early-type spiral galaxy using SCORPIO long-slit data. A minor merging event is also a most likely origin for  the full-size  gaseous discs  rotating  in the opposite direction to the stellar ones in NGC2551 and NGC5631  lenticular galaxies (Sil'chenko et al, 2009).

The SCORPIO advantages in the spectroscopy of regions with a low surface brightness are illustrated in the papers by Zasov et al. (2008) about stellar kinematics of the discs in S0-Sa galaxies and  by  Baes et al. (2007), where stellar population age and metallicity distributions in the sample of elliptical galaxies were estimated up to distances  of 3 effective radii. The main important conclusion of this work is the absence  of a single power law for the metallicity gradient that is  inconsistent with the origin of the elliptical galaxies by a major merger.

\subsectionb{3.4}{Nearby AGN}

Together with data taken from some other instruments the SCORPIO spectra were involved in  long-term monitoring of   H$\alpha$ and H$\beta$ lines variations of the active galactic nucleus of NGC4151 (Shapovalova et al.  2008) and 3C390.3 (Popovi{\'c} et al.  2011). The main aim is a study of their `central engine'  including Broad Line Region (BLR). The geometry of the BLR of 3C390.3 seems to be very complex, and inflows/outflows may be present, but the disc-like BLR  is the dominant emitter.

Recently,  Afanasiev et al. (2011) presented the results of  spectropolarimetric observations for a sample of 15  active galactic nuclei. The magnetic field strengths and radial distributions in an accretion disc around a supermassive black hole were evaluated within the framework of traditional accretion disc models.

The large-scale environment of active nuclei was also investigated in numerous papers based on the data collected in the  FPI mode. Smirnova et al. (2007) presented the analysis of global ionized gas kinematics in the disc of  Mrk 533. In this galaxy the  non-circular ionized gas motions at the distance of  $r<2.5$ kpc are associated with an  outflow triggered by the nuclear radio jet intrusion in an ambient medium. A very complicated  combination of the region with different ionization and kinematics properties was found in  Mrk 344 (Smirnova \& Moiseev 2010).  The most unusual feature is a large-scale cavern   filled with a low-density ionized gas. This region seems  to be the place where the remnants of a disrupted companion have recently penetrated through the gaseous disc of the main
galaxy.

\subsectionb{3.5}{Distant Objects}

SCORPIO shows a good advantage in the spectral identification of the extragalactic radio sources in a wide range of optical magnitudes  up to $m_r=23-24^m$. See, for example, the   classifications, optical identifications  and spectral redshifts  for  the different samples of radio sources presented by Amirkhanyan et al.  (2004) and  Afanasiev et al. (2003). Some interesting objects were discovered. For instance,  Amirkhanyan \& Mikhailov (2006) found a very radio-loud QSO at $z = 4.06$. Recently Parijskij et al. (2010) presented the results of   spectroscopy of 71 radio galaxies and QSO with steep and ultra-steep spectra. 	

SCORPIO follow-up spectroscopy makes a  significant contribution  to the  systematic searches for wide separation  gravitational lens systems in the framework of CAmbridge Sloan Survey Of Wide ARcs in the skY (CASSOWARY). The most  beautiful object (in our view) was the discovery of the Cosmic Horseshoe (CASSOWARY \#1) an almost complete Einstein ring of the diameter of  $10''$ around a giant luminous red galaxy at the $z=0.444$ (Belokurov et al. 2007). The source is a star-forming galaxy that has a $z=$2.379. This gravlens  has a large magnification factor ($25-35$) which allows Quider et al. (2009) to study from VLT spectroscopy  the metallicity  and starformation properties in the source galaxies with the quality that   is currently unfeasible for unlensed galaxies at $z\approx 2-3$.

\sectionb{4}{NEW PERSPECTIVIES}

During its ten years of operation the SCORPIO has been repeatedly upgraded  and improved. Unfortunately, opportunities for further upgrading  have been  exhausted. Also, a new optical scheme was necessary for spectral observations   with  large format CCD detector.  Therefore SAO RAS began manufacturing a new multi-mode spectrograph with enhanced capabilities. The main novelty of the SCORPIO-2 versus its with previous version  are as follows (see also  Table~1):

\begin{itemize}
\item The value of off-axis optical aberration are significantly (by half) decreased.

\item The device is specially designed to work under remote control from the Institute building (under the mountain where the  telescope is sited). The number of exchangable elements installed simultaneously in the device is significantly increased.
\item The opportunities for polarimetry (spectra and images)  are greatly expanded.
\item The new multi-mode focal reducer includes an Integral-field unit (IFU) based on the combination of small lenses with optical fibers.  This scheme was offered by Georg Court\'{e}s (1982) and it was first implemented in the two  generations of the MultiPupil Fibers Spectrographs (MPFS) at the 6-m telescope (Afanasiev et al.  1990, 2001). Now this type of IFUs is  widely used  in middle- and large-size telescopes. The SCORPIO-2/IFU $18\times18''$ field-of view  is divided by  square lenses array with a scale of $0.75''$ per lens.  Behind each lens an optical fibre is located whose other end is packed into two pseudo-slits in the  spectrograph entrance.
\end{itemize}

The first test observations by  the 6-m telescope were carried out in  June, 2010. Some electronic  and mechanical parts (integral-field and  multi-slit units) are still under construction. We are confident that the commissioning of SCORPIO-2 will significantly enhance the abilities  of the 6-m telescope in the study of different objects in our Galaxy as well as in extragalactic scales.

\begin{table}[!t]
\begin{center}
\vbox{\footnotesize\tabcolsep=3pt
\parbox[c]{124mm}{\baselineskip=10pt
{\smallbf\ \ Table 1.}{\small\
Comparision of the 6-m telescope old and new facilities\lstrut}}
\begin{tabular}{l|l|l}
\hline
                           &     SCORPIO              & SCOPIO-2 \\
\hline
Detector                   & EEV~42-40, $2K\times2K$ & E2V~42-90, $2K\times4.6K$ \\
\hline
Direct imaging: & & \\
Max. filters positions & 10 (in two wheels)  &  27 (in three wheels) \\
Field-of-view & 6.1 arcmin  & 6.1 arcmin \\
\hline
Long-slit spectroscopy & set of slits with fixed width ($0.5-2'')$;& variable slit width ($0-20'')$ ;\\
                       & single VPHG position                         & wheel with 9 grating holders \\
\hline
FPI                    & Common carriage with grating holder    & independent holder   \\
\hline
Multi-slit unit        & 16  slits in $6\times3$ arcmin field-of-view &  16  slits in $6\times4$ arcmin field-of-view  \\
\hline
Integral-field unit    & --                 & $24\times24$ lenslet, $0.75''/$lens \\
\hline
Polarymetry            & \parbox{4.5 cm}{ Savar plates, rotated in two positions} &  \parbox{4.8 cm}{Single and double Wollaston prisms;\\apochromatic phase plates $\lambda/2$, $\lambda/4$;\\ rotated analyser} \\
\hline
\end{tabular}
}
\end{center}
\vskip-6mm
\end{table}

\thanks{This work was supported by the Russian Foundation for Basic Research (project
no.~09-02-00870) and by the Russian Federal Program `Kadry' (contract no.~14.740.11.0800). We are grateful to Olga Smirnova for her help in the text preparation. AVM is also grateful to the Dynasty Fund and the 8th SCLCA Organizing Committe  for their financial support.}

\References

\refb Abolmasov P., Maryeva O., Burenkov A. N. 2010, AN, 331, 412

\refb Abolmasov P., Moiseev A. 2008, MexRevA\&A, 44, 301

\refb Afanasiev  V. L., Borisov N. V., Gnedin Yu. N.  2011, AstL, 37, 302

\refb Afanasiev  V. L., Dodonov  S. N., Sil'chenko O. K., Vlasyuk V. V. 1990, Preprint Spec. Astrophys. Obs., N54

\refb Afanasiev V. L., Dodonov S. N., Moiseev A. V. 2001, in {\it Stellar Dynamics: From Classic to Modern}, eds.  L.P. Ossipkov \& I.I. Nikiforov, Saint Petersburg, 103

\refb Afanasiev  V. L., Dodonov  S. N., Moiseev  A. V. et al. 2003, ARep, 47, 458

\refb Afanasiev  V. L., Gazhur E. B., Zhelenkov S. R., Moiseev A. V.  2005, Bull. Spec. Astrophys. Obs. 58, 90

\refb Afanasiev  V. L.,  Kalenichenko V.V.,  Karachentsev I.D.  2007, AstBu, 62, 301

\refb Afanasiev  V. L., Moiseev A. V.  2005, AstL, 31, 194

\refb  Amirkhanyan  V. R, Afanas'ev V. L., Dodonov S. N., Moiseev A. V., Mikhailov V. P.  2004, AstL, 30, 834

\refb Amirkhanyan V. R., Mikhailov V. P. 2006, Astrophysics, 49, 184

\refb Baes M, Sil'chenko O. K., Moiseev A. V., Manakova E. A. 2007, A\&A, 467, 991

\refb Belokurov V. , Evans N.W., Moiseev A. et al. 2007, ApJ, 671, 9L

\refb Brosch N., Kniazev A., Moiseev A., Pustilnik S. 2010, MNRAS, 401, 2067

\refb Buzzoni, B., Delabre B., Dekker H. et al.  1984, ESO Messenger, Dec. 1984, 9

\refb Chilingarian I,  Cayatte V., Revaz Y. et al. 2009, Science, 326, 1379

\refb Fabrika S., Sholukhova O., Valeev A., Hornoch K., Pietsch W. 2009, ATel, 2239, 1

\refb Korsun P. P., Ivanova O. V., Afanasiev V. L.  2008, Icarus, 198, 465

\refb Korsun P. P., Kulyk I. V., Ivanova O. V. et al.  2010, Icarus, 210, 916

\refb Court\'{e}s G.  1960, Ann. d’Astrophysics, 23, 115

\refb Court\'{e}s G.  1982, in {\it Instrumentation for Astronomy with Large Optical Telescopes}, eds. C. M. Humphries, Astrophysics and Space Science Library,  92, 123

\refb Gvaramadze V. V.,  Fabrika S., Hamann W.-R., et al. 2009, MNRAS, 400, 524

\refb Karachentsev I. D., Kaisin S. S. 2010, AJ, 140, 1241

\refb Lozinskaya T. A., Komarova V. N., Moiseev A.V., Blinikov S.I. 2005, AstL, 31, 243

\refb  Lozinskaya T. A., Moiseev A. V.   2007, MNRAS, 381, 26L

\refb  Lozinskaya T. A.,  Moiseev A. V.,   Podorvanyuk N.Yu.  2003, AstL, 29, 77

\refb Makarova L., Koleva M., Makarov D., Prugniel P. 2010, MNRAS, 406, 1152

\refb  Mart\'{\i}nez-Delgado I., Tenorio-Tagle  G., Mu{\~n}oz-Tu{\~n}{\'o}n  C., Moiseev A.~V., Cair{\'o}s L.~M. 2007, AJ, 133, 2892

\refb Moiseev A.V.  2002, Bull. Spec. Astrophys. Obs., 54, 74

\refb Moiseev A. V. 2007, in {\it Science Perspectives for 3D Spectroscopy}, eds.  M. Kissler-Patig, M.M. Roth \& J.R. Walsh, ESO Astrophysics Symposia,  105

\refb  Moiseev A. V. 2010, in  {\it  A Universe of dwarf galaxies}, eds. M. Koleva, Ph. Prugniel \& I. Vauglin,  arXiv:1009.2519

\refb Moiseev A. V.,  Bizyaev D. V. 2009,  New Astronomy Reviews, 53, 169

\refb Moiseev A. V., Egorov O. V. 2008, AstBu, 63, 181

\refb Moiseev A. V.,  Pustilnik S.A.,  Kniazev A.Y. 2010, MNRAS, 405, 2453

\refb Moskvitin  A. S., Fatkhullin  T. A., Sokolov V. V. et al. 2010, AstBu, 65, 230

\refb Movsessian T. A., Magakian T. Yu., Bally J. et al. 2007, A\&A, 470, 605

\refb Movsessian T. A., Magakian T. Yu., Moiseev A. V.,   Smith M.D. 2009, A\&A, 508, 773

\refb  Parijskij Y. N., Kopylov A. I., Temirova A. V. et al. 2010, ARep, 54, 675

\refb Popovi{\'c} L. {\v C}.,  Shapovalova A.I.,  Ili{\'c} D. et al. 2011, A\&A, 528, A130

\refb  Pustilnik S. A.,  Tepliakova A.L.,  Kniazev A.Y.,  Burenkov A.N.  2008, MNRAS, 388, 24L

\refb  Pustilnik  S. A., Tepliakova  A. L., Kniazev  A. Y., Martin  J.-M., Burenkov   A. N. 2010, MNRAS, 401, 333

\refb Quider A. M., Pettini M., Shapley A. E., Steidel C. C. 2009, MNRAS, 398, 1263

\refb Sil'chenko O. K.,  Moiseev A.V.,  Afanasiev V.L. 2009, ApJ, 694, 1550

\refb Sil'chenko O., Chilingarian I., Sotnikova N., Afanasiev V., 2011, MNRAS, in press; arXiv:1103.1692

\refb Smirnova  A. A., Gavrilovi{\'c} N., Moiseev A. V et al. 2007, MNRAS, 377, 480

\refb Smirnova A., Moiseev A.  2010, MNRAS, 401, 307

\refb Shapovalova A. I.,  Popovi{\'c} L. {\v C}.,  Collin S. et al. 2008, A\&A, 486, 99

\refb Valeev A. F., Sholukhova O., Fabrika S. 2009, MNRAS, 396, L21

\refb Zasov A. V., Moiseev  A. V., Khoperskov A. V., Sidorova E. A. 2008, ARep, 52, 79
\end{document}